\documentclass[12pt]{article}
\usepackage{float}

\usepackage[superscript,biblabel]{cite} 

\usepackage{times}
\usepackage{graphicx}
\usepackage{siunitx}
\usepackage{subcaption}
\usepackage{textcomp} 
\usepackage{lineno} 
\usepackage[hidelinks]{hyperref}
\usepackage{microtype} 
\usepackage{amsmath}
\usepackage{comment}

\usepackage{lineno}

\newenvironment{sciabstract}{%
\begin{quote} \bf}
{\end{quote}}

\title{Amplitude- and frequency-modulated combs from an actively locked metasurface external-cavity laser}

\author
{Marco Raffa$^{1\ast}$, Jordane Bloomfield$^2$, Yu Wu$^2$,\\
Sadhvikas J. Addamane$^3$, Alexander Dikopoltsev$^{1}$, J\'er\^ome Faist$^{1}$,\\
Benjamin S. Williams$^{2}$, Giacomo Scalari$^{1\ast}$\\
\\
\normalsize{$^{1}$Institute for Quantum Electronics, ETH Z{\"u}rich, 8093 Z{\"u}rich, Switzerland}\\
\normalsize{$^{2}$University of California Los Angeles, Department of Electrical and Computer Engineering,}\\
\normalsize{Los Angeles, CA 90095, United States}\\
\normalsize{$^{3}$Sandia National Laboratories, Center of Integrated Nanotechnologies,}\\
\normalsize{MS 1303, Albuquerque, New Mexico 87185, United States}\\
\normalsize{$^\ast$To whom correspondence should be addressed;}\\
\normalsize{E-mail: mraffa@phys.ethz.ch, scalari@phys.ethz.ch}
}

\date{}

\begin{document} 




\maketitle 

\begin{sciabstract}
Optical frequency combs are key components  of several photonics applications including spectroscopy, communications, and ultrafast photonics. A central challenge in frequency-comb photonics is to develop sources whose operating state can be precisely controlled and adapted to different application needs. We introduce frequency comb functionality to a THz metasurface vertical-external-cavity-surface-emitting laser (VECSEL), combining its characteristic high output power and excellent beam quality with a reconfigurable comb output. The source exhibits reversible switching between actively mode-locked 3.5 ps-long pulses and stable frequency-modulated quantum walk comb states. The flexible control of the intermodal phase relation is achieved through careful dispersion engineering via a Gires-Tournois interferometer (GTI) output-coupler combined with resonant RF bias modulation of the metasurface. These results pave the way for on-demand comb control in the THz range and provide a versatile strategy that could be extended to other semiconductor frequency-comb platforms and wavelength ranges.

    
\end{sciabstract}
\begin{figure}
  \centering
  \includegraphics[width=\textwidth]{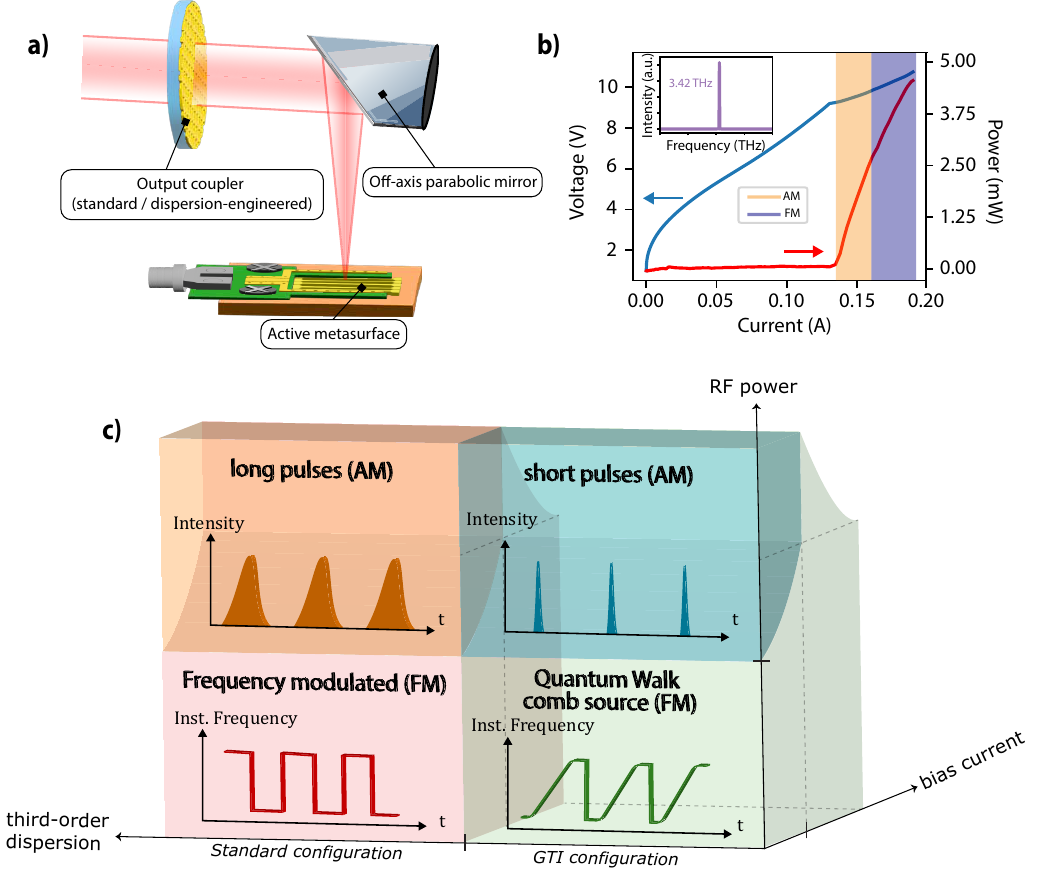}
  \caption{\footnotesize \textbf{VECSEL schematic and operating regimes.} \textbf{a)} Illustration of the VECSEL platform operating in standard configuration. \textbf{b)} CW LIV curve of the VECSEL at 25 K in free-running. Spectrum with single mode emission centered at 3.42 THz (inset). \textbf{c)} Switchable VECSEL operating regimes as function of the bias point, RF modulation power and dispersion engineering. Pink region: third-order dominated frequency modulated regime at low RF modulation power ($\leq+20$ dBm at the source) in standard configuration. Green region: Quantum Walk comb frequency modulated regime at low RF modulation power adding group delay dispersion through GTI configuration. Orange region: long pulses in active mode-locking regime at low bias, under strong RF modulation power ($\geq+20$ dBm at the source) in standard configuration. Blue region: short pulses in active mode-locking regime at low bias, under strong RF modulation power with dispersion engineering.} 
  \label{fig:fig1}
\end{figure}

\section*{Main}\label{main}
\subsection*{Introduction}
Terahertz (THz) sources are an enabling technology for applications ranging from spectroscopy and imaging \cite{liu2007terahertz,pickwell2006biomedical} to high-capacity wireless communications\cite{koenig2013wireless}. Considerable effort has been devoted to developing compact semiconductor-based THz emitters that combine high output power, low phase noise, wide spectral coverage, tunability, and high beam quality\cite{senica2026continuously,senica2022planarized}. In this context, both single-mode and multimode frequency-comb THz sources are of particular interest. Nonlinear approaches such as optical rectification, difference frequency generation  and photomixing \cite{schneider2006generation,THz_photomixer,THzDFGNatPhot_Belkin_2026} rely on indirect THz generation, whereas quantum cascade lasers (QCLs) enable direct  frequency-comb generation in mid-IR\cite{Hugi:2012ep} and THz\cite{burghoff_terahertz_2014,Roesch2014,garrasi_high_2019}.

Several strategies have been developed to improve THz comb performance, including passive outcoupling antennas for low-divergence emission\cite{senica2023broadband}  and dispersion engineering for broadband emission in metal–metal THz QCLs\cite{micheletti_terahertz_2023,bachmann2016dispersion,xu_design_2016,Roy_Burghoff_OPtica_24}. Fabry–Pérot ridge devices, because of standing-wave formation, suffer from spatial hole burning (SHB), which promotes multimode formation. Four-wave mixing (FWM) and RF modulation can lock these modes to form an optical frequency comb, but cavity dispersion—including higher-order contributions such as third-order dispersion—induces non-uniform intermodal phase evolution, limiting phase synchronization and preventing robust comb formation. Consequently, the accessible dynamical regimes exhibit reduced stability and reproducibility.\cite{Opacak:24_high_order_dispersion} Recently, the quantum walk (QW) laser  \cite{heckelmann_quantum_2023,Dikopoltsev_theoryQW_nanophotonic} has been introduced for comb generation. In the absence of SHB and under RF modulation, fast gain dynamics, much faster than the cavity round-trip time, combined with quadratic dispersion enable coherent coupling of modes in synthetic frequency space, resulting in a stable, broadband frequency-modulated (FM) comb state. The concept has proven broadly transferable across platforms and wavelength ranges, with demonstrations in mid-infrared \cite{heckelmann_quantum_2023}  and THz quantum cascade ring lasers \cite{DigiorgioTHzQwalk2026}, ring external cavities at telecommunications wavelengths using interband semiconductor optical amplifiers\cite{marzban2026quantum} and, more recently, in experiments exploiting synthetic-frequency-space dynamics for programmable spectral shaping, broadband comb spectroscopy, and collective light-flow control.\cite{Piciocchi_shaping,heckelmann2026broadband,dikopoltsev2025collective}. To suppress spatial hole burning (SHB), ring cavities are typically employed; however, their dispersion is fixed after fabrication, leaving no possibility for further compensation engineering, especially in fully integrated implementations, which often suffer from reduced output power and degraded beam quality. Alternative approaches based on racetrack resonators, bus waveguides, or vertically coupled waveguides have been developed to recover output-power performance.\cite{letsou2026high,Cargioli_QW_dual_waveguide}.

More recently, THz quantum-cascade metasurface vertical-external-cavity surface-emitting lasers (VECSELs) \cite{Tropper2004,Maas2007} — consisting of an amplifying reflector array of metal–metal ridge antennas and paired with an output coupler (OC) mirror (Fig.\ref{fig:fig1}a)— have emerged as a promising platform for spectroscopy, demonstrating high-power (watt level in pulsed mode, milliwatt in continuous wave) single-mode operation with excellent beam quality\cite{curwen_terahertz_2018,metasurface_VECSEL_Xu_2015}. 
Unlike monolithic THz QCLs, in a QC-VECSEL the longitudinal modes primarily reside within vacuum except where they overlap with the QC gain material within the metasurface, that is highly subwavelength ($d/\lambda\approx 0.3 $) in the longitudinal direction. Interaction of each longitudinal mode with the active material is mediated through the same metasurface resonance and spatially samples the material gain in the same way. This leads to effective gain clamping at lasing threshold and stable single-mode operation closer to the gain peak rather than spontaneous comb formation, even when the longitudinal mode spacing is around 100 times smaller than the gain bandwidth [Fig.\ref{fig:fig1}b]\cite{metasurface_VECSEL_long_cavity_Wu_2022}. Such a SHB-free system would be expected to offer access to the aforementioned quantum walk comb regimes without the need for a ring cavity. 
Furthermore, the roundtrip time in a QC-VECSEL can be readily adjusted over large intervals by modifying the cavity dimensions, either through static design, or dynamically using a piezoelectric actuator enabling high-resolution mapping of absorption lines \cite{curwen2019broadband,VECSEL_absorption_spectroscopy_Morag_2026}.
Although strong RF modulation can drive THz VECSELs into rich multimode regimes \cite{Wu_RF_harmonic_subharmonic_VECSEL}, transforming this behavior into a controllable frequency-comb platform requires direct control of the intermodal phase dynamics and a clear understanding of how cavity dispersion governs the temporal structure of the emission.

To overcome these limitations, we propose and implement a dispersion-compensation scheme based on a Gires–Tournois interferometer (GTI) output coupler\cite{golubovic2000double}. By tailoring the cavity dispersion, the GTI enables controlled adjustment of lasing regimes and intermodal phase relationships. This allows deterministic switching between active mode-locked pulse emission and the quantum walk (QW) comb regime in the THz VECSEL platform depending on the RF modulation power and bias point [Fig.\ref{fig:fig1}c]. 
The THz metasurface VECSEL is particularly well suited for implementing the quantum walk comb regime because it is intrinsically free from SHB, has high power and good beam quality and is single mode by design. In this work, we focus on implementing the dispersion-compensated scheme and assessing the coherence of the resulting lasing emission.
\begin{figure}[h]
    \centering
    \includegraphics[width=\linewidth]{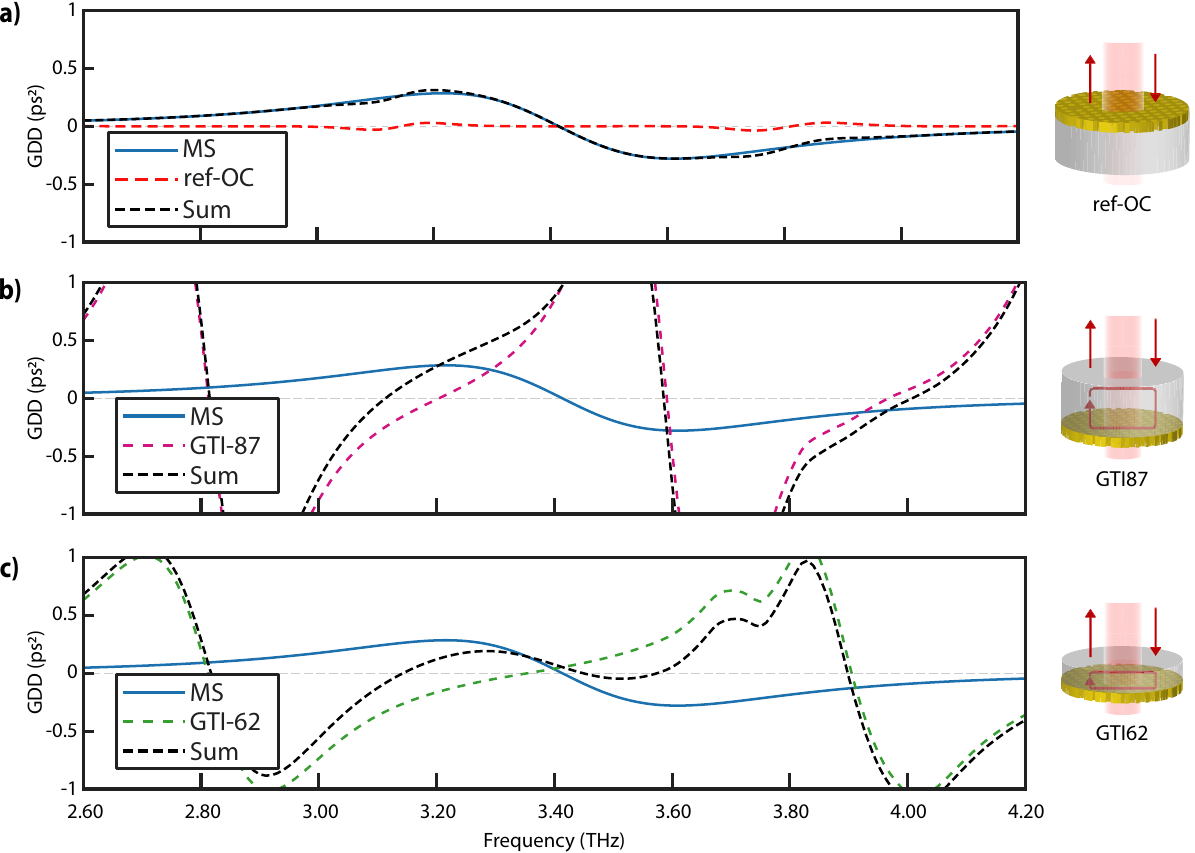}
    \caption{\footnotesize \textbf{Calculated GDD of the active metasurface, each output couplers, and their sum over one round-trip.}}
    \label{fig:OCs_GDD}
\end{figure}

\subsection*{Results}\label{results}
The QC-metasurface consists of an amplifying reflectarray of metal-metal ridge antennas loaded with 10$\mu$m-thick GaAs/Al\textsubscript{0.15}Ga\textsubscript{0.85}As heterostructure active region \cite{Wu_RF_harmonic_subharmonic_VECSEL}. The metasurface is spatially homogeneous, and each ridge is 11.8 $\mu$m wide with a 41.7 $\mu$m period, which centers the Lorentizan resonance at 3.4 THz with a radiative quality factor of $\approx$ 6.
The external cavity is 24 mm long, and is realized by pairing the active metasurface to the output coupler through an off-axis paraboloid (OAP) mirror (see  Fig.\ref{fig:fig1}a).

To evaluate the comb dynamics, the modes coupled by RF modulation can be described as a lattice in the frequency domain, where each cavity mode is mapped onto a site of a synthetic lattice \cite{Dikopoltsev_theoryQW_nanophotonic, Piciocchi:26}. In this synthetic space, the detuning $\Delta\Omega$ from the round-trip cavity resonant frequency $\Omega$\textsubscript{res}, the gain curvature $D_g$ associated with the finite gain bandwidth, the dispersion terms $D_i$, and the modulation depth $M$ of the RF signal contribute to the effective complex potential. Considering a single mode emission centered at frequency $\omega_0$, the modulation depth $M$ acts as a coupling mechanism between neighboring frequency sites. The group delay dispersion (GDD) sets a boundary to the coupling mechanism through a parabolic potential $D_2$, while the gain curvature $D_g$ acts as a dissipative mechanism with an imaginary parabolic potential. Higher-order dispersion contributions, such as the third-order dispersion (TOD), increase anharmonicity in the effective potential and counteract the confinement of the modes in the parabolic landscape. The detuning $\Delta\Omega$ from the cavity resonant frequency $\Omega$\textsubscript{res} tilts the parabolic effective potential resulting in a shift of the equilibrium point of the comb expansion.

Considering only the real part, the confinement potential is:
\begin{equation}
U(\omega)=-M +\frac{2\pi\Delta\Omega}{\Omega_{res}^2}(\omega-\omega_0)+D_2(w_0)(\omega-\omega_0)^2+D_3(w_0)(\omega-\omega_0)^3+...
\label{eff_pot}
\end{equation}

Unlike monolithic devices in which the optical field is mainly contained inside the active material, in the QC-VECSEL platform the cavity modes propagate mainly in the empty cavity space. Therefore the chromatic dispersion primarily accumulates on a per-reflection basis from both the metasurface and the output coupler. Ordinarily an  
metallic mesh output coupler in standard configuration (THz intracavity field incident on the mesh side termed \textit{ref-OC}) contributes negligible dispersion.
On the other hand, the uniform single-resonator metasurface used here exhibits a Lorentzian-type resonance characteristic, in which the reflection phase changes by nearly 2$\pi$ phase shift across its spectral bandwidth; this impacts the group delay, group delay dispersion (GDD), and third-order dispersion (TOD). Indeed, it is likely that the dispersive contributions of the intersubband transitions themselves are comparatively minor \cite{Metasurface_TDS_Shen_2021}. Furthermore, for the standard output coupler configuration, in which the sub-wavelength inductive metal-mesh faces inward to the cavity, the output coupler also contributes negligible dispersion\cite{Wu_RF_harmonic_subharmonic_VECSEL}. A summary of the metasurface and output couplers dispersion contributions is given in Fig.\ref{fig:OCs_GDD}. 
The strong dispersion produced by the metasurface is potentially problematic, as the mapping of the QW comb regime to a quantum harmonic oscillator is no longer valid in the presence of a high third-order dispersion term \cite{Dikopoltsev_theoryQW_nanophotonic}.

\begin{figure}
    \centering
    \includegraphics[width=1\linewidth]{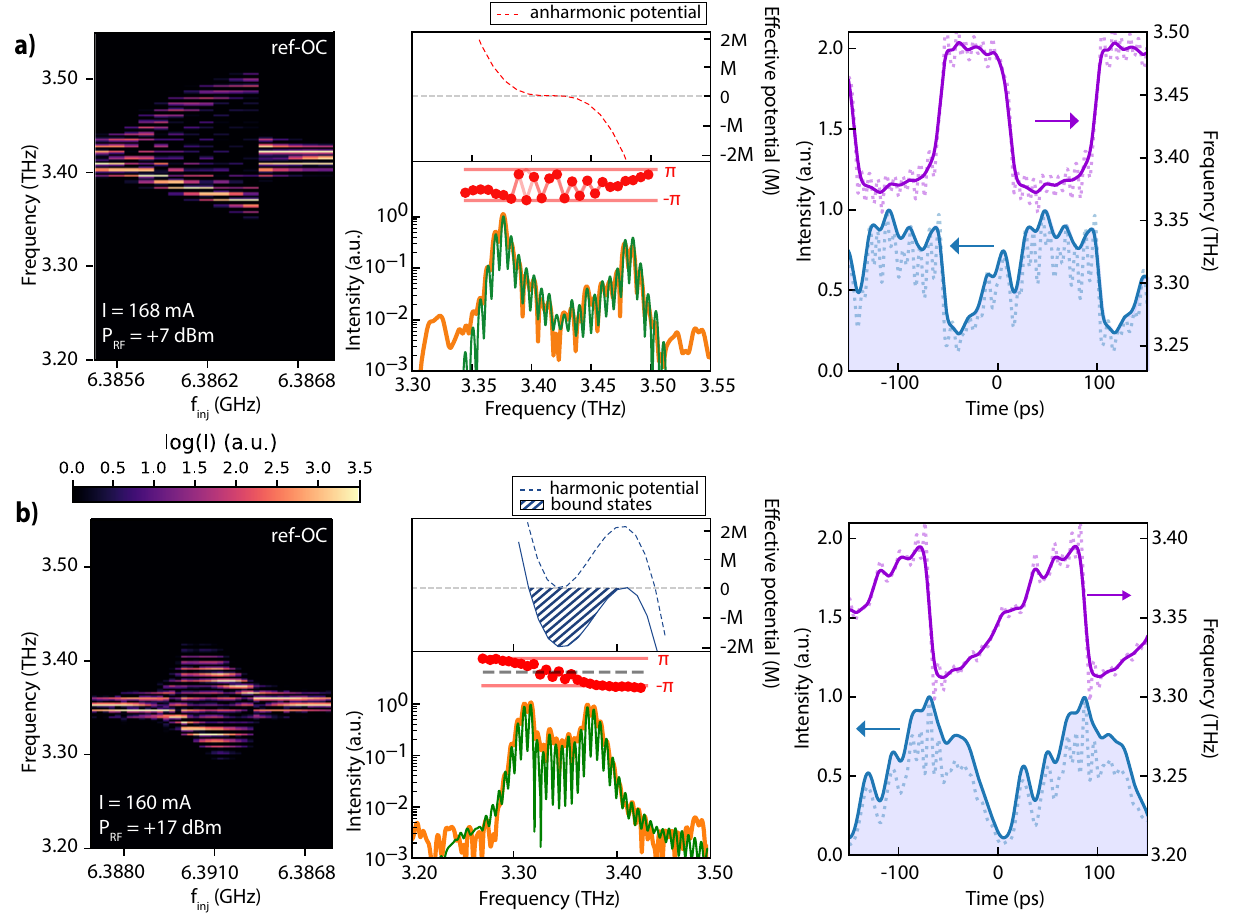}
    \caption{\footnotesize \textbf{RF modulation at low injection power for VECSEL in standard configuration.} \textbf{a)} Left panel: spectral map around fundamental roundtrip frequency with \textit{ref-OC} dominated by third-order dispersion. Central panel: spectral product (green), SWIFT spectrum (orange), intermodal phase differences close to resonance and effective potential in presence of third-order dispersion. Right panel: intensity and double discontinuity of the instantaneous frequency over time in one cavity round-trip close to resonance, reconstructed [dots] and smoothed [solid trace]. \textbf{b)} Left panel: spectral map around fundamental cavity frequency with \textit{ref-OC} after exploiting bistable single mode frequency-shift. Central panel: spectral product (green), SWIFT spectrum (orange), intermodal phase differences close to resonance and effective potential. Right panel: intensity and instantaneous frequency at resonance with half-cosine behavior and a single discontinuity point, reconstructed [dots] and smoothed [solid trace].}
    \label{fig:QW_ref_OC}
\end{figure}

In the first experimental characterization, the QC-VECSEL in standard configuration (\textit{ref-OC}) was cooled down to 25 K and then operated in continuous-wave (CW) mode. 
THz emission spectra are measured as the RF modulation frequency is swept near the round-trip frequency; the resulting spectral map of THz spectra vs. RF frequency is plotted in 
Fig.\ref{fig:QW_ref_OC}a (left). RF injection around the harmonic roundtrip frequency $2 \times f_{rep}$ and around the subharmonic roundtrip frequency $f_{rep}\div 2$ are also investigated [Fig.S1a].
In all three injection conditions reported, the lasing emission broadens as it approaches the resonant RF frequency, exhibiting distinctive lobes at the edges of the spectral envelope.
As expected, the optical modes are locked at the fundamental frequency when signals close to the round-trip frequency are injected into the device.
The coherence of the broadened state is assessed using the Shifted Wave Interference Fourier Transform (SWIFT) spectroscopy technique \cite{Burghoff:2015_SWIFTS,senica2022planarized}. The multimode laser output passes through a Fourier Transform Interferometer (FTIR) and is detected by a fast detector ($\sim$ 0.5 GHz to 40 GHz, a Schottky diode from VDI in this case), where the optical beating between adjacent lasing modes is demodulated at the roundtrip frequency. The spectrum product, measured with a slow deuterated triglycine sulfate (DTGS) detector, and the SWIFT spectrum are plotted together in Fig.\ref{fig:QW_ref_OC}a (central); they show good agreement which is proof of a high degree of coherence in the lasing emission. The intermodal phase difference was retrieved to reconstruct the time profile of the comb emission. In this regime of moderate RF modulation power, pulse formation is suppressed by the fast gain dynamics, resulting in a quasi-continuous wave output and double discontinuity points in instantaneous frequency, reported in Fig.\ref{fig:QW_ref_OC}a (right). 
The abrupt discontinuities observed in the spectral maps as the RF modulation frequency is increased are associated with the high third-order dispersion of the QC-VECSEL platform in its standard configuration, as shown in Fig. S2. The central panel of Fig.\ref{fig:QW_ref_OC}a shows that the effective confinement potential from Eq. \eqref{eff_pot}, retrieved using finite-element-method simulations of the group-delay dispersion (GDD) of both the active metasurface medium and the output coupler, is highly anharmonic. As a result, modes in the synthetic space are not confined by a parabolic potential across the RF modulation-frequency sweep. Although comb operation is preserved, as confirmed by the SWIFT measurements, the absence of a continuous spectral evolution with RF frequency, together with the discontinuity of the instantaneous frequency at half the round-trip time, indicates that an anharmonic effective potential leads to a more general FM comb state in the presence of high third-order dispersion, rather than to the specific quantum walk comb laser regime.\cite{Opacak:24_high_order_dispersion,Dikopoltsev_theoryQW_nanophotonic,Cargioli_QW_dual_waveguide}

\begin{figure}[h]
    \centering
    \includegraphics[width=\linewidth]{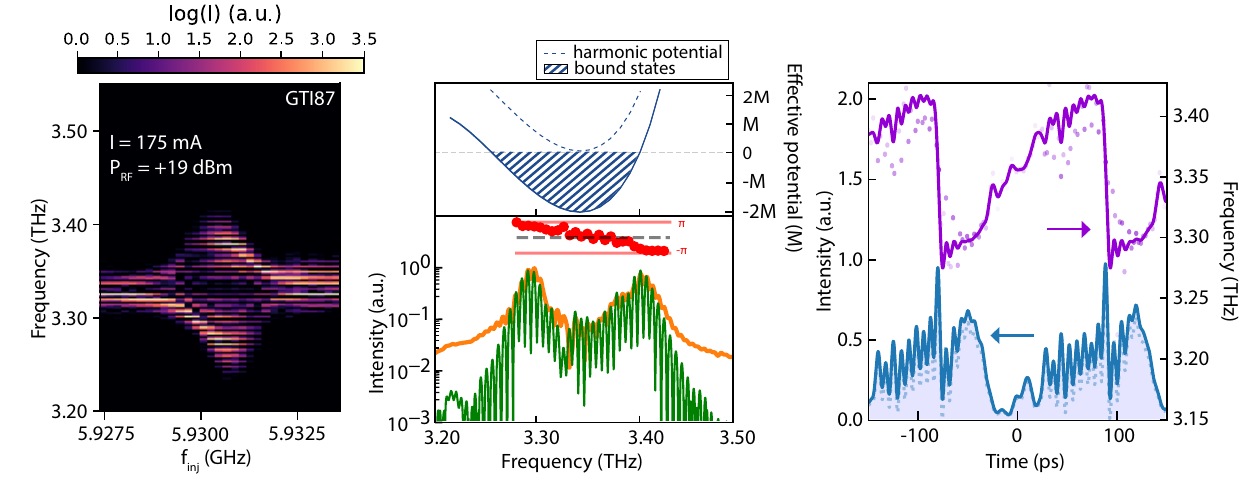}
    \caption{\footnotesize \textbf{RF modulation at low injection power for VECSEL with GTI87.} Left panel: broader bandwidth spectral map at fundamental RF modulation. \textbf{b)} Central panel: spectral product (green), SWIFT spectrum (orange), intermodal phase differences close to resonance and restored harmonic potential. Right panel: intensity and instantaneous frequency at resonance with half-cosine behavior and a single discontinuity point, reconstructed [dots] and smoothed [solid trace].}
    \label{fig:fig3}
\end{figure}

As the injected power at the RF resonant frequency is increased, an abrupt red-shift jump in the multimode THz emission is observed. However, when the RF modulation is subsequently reduced and switched off, the system does not return to the original single mode emission centered at 3.42 THz, resulting in a hysteretic relationship between the optical frequency and the RF power (more details in Supplementary). The RF modulation acts as a nonlinear perturbation that drives the QC-VECSEL across a static modal bistability boundary; once this boundary is crossed, the new CW lasing mode becomes self-stabilized and persists even in the absence of RF modulation. This procedure enables single-mode emission at 3.35 THz, in a spectral region less affected by third-order dispersion, corresponding to the “walker” in a harmonic potential in the continuous-time quantum walk comb laser dynamics.

Starting from the single-mode state achieved through the previous procedure, RF modulation is applied, yielding the continuous spectral map with distinct on- and off-resonant lasing regimes shown in Fig.\ref{fig:QW_ref_OC}b (left). The off-resonance and slightly shifted-from-resonance spectra are investigated by SWIFT spectroscopy, and the corresponding temporal dynamics are retrieved in Fig. S4. The comb in the resonant state is shown in Fig.\ref{fig:QW_ref_OC}b (central) together with a parabolic confinement potential, from Eq. \eqref{eff_pot}, extending over a limited THz bandwidth. Unlike in the previously described general FM regime, the on-resonance instantaneous frequency follows a half-cosine behavior with a single discontinuity point, Fig.\ref{fig:QW_ref_OC}b (right), as predicted by QW comb laser theory. \cite{Dikopoltsev_theoryQW_nanophotonic}. This highlights the importance of a well-defined parabolic cavity potential for efficiently locking the emission into a QW comb state without spectral jumps. Taking advantage of the bistable behavior of the VECSEL cavity, spectral maps are also successfully recorded around harmonic and subharmonic cavity frequencies,Fig.S3 ,making it possible to extend QW comb operation beyond dispersive ring cavities.

Next, we replaced the conventional \textit{ref-OC} output coupler with a Gires-Tournois interferometer (GTI) output coupler (\textit{GTI87}),
which introduces a frequency-dependent phase shift when the intracavity THz field impinges on the quartz side of the output coupler rather than on the metallic mesh side (see Methods). 
The GTI introduced additional dispersion such that an effective  harmonic potential from Eq. \eqref{eff_pot} is obtained over a wider bandwidth (see Fig.\ref{fig:fig3} (central)) \cite{villares2016dispersion}.
There are two main results. First, the central lasing frequency is shifted down slightly to near 3.32 THz due to the change in the OC reflectance spectrum. Second, a stable QW comb spectral map with larger bandwidth was obtained, without the need to rely on the bistable dynamics associated with RF injection to reach this operating state, Fig.\ref{fig:fig3} (left). Intensity and instantaneous frequency at resonance are retrieved with SWIFT spectroscopy and plotted in Fig. \ref{fig:fig3} (right).

\begin{figure}[!]
    \centering
    \includegraphics[width=\linewidth]{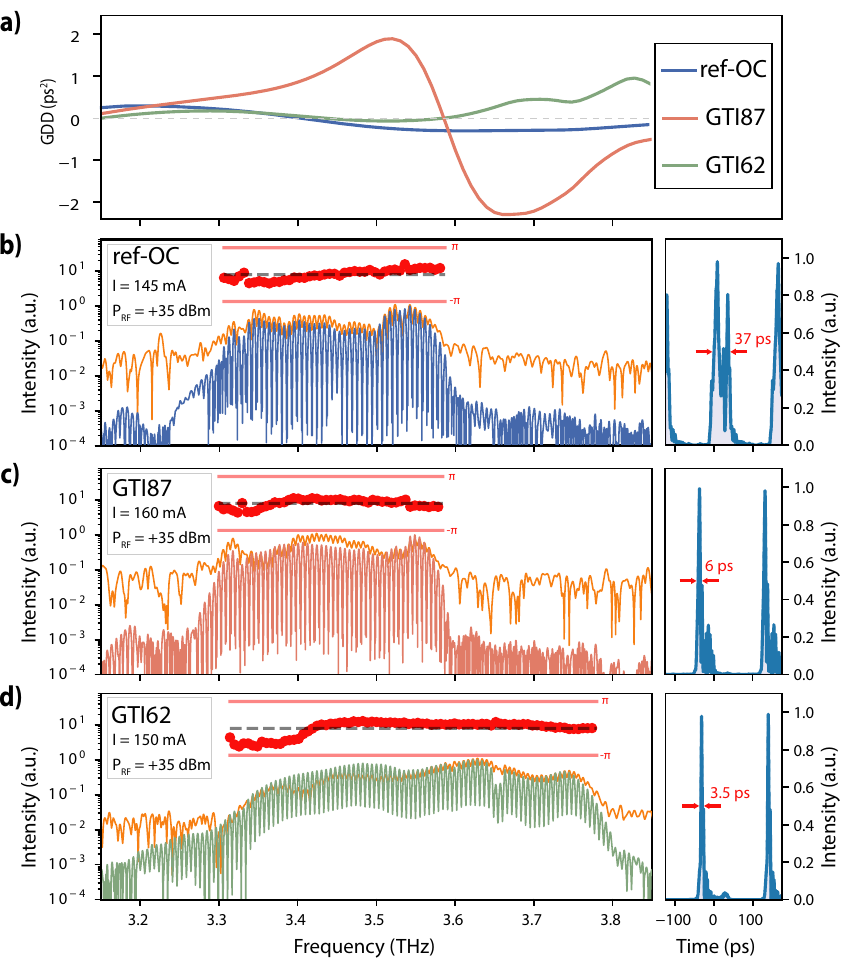}
    \caption{\footnotesize \textbf{Active mode-locked regimes with pulse formation.} Spectral product (color-coded as defined in the legend), SWIFT spectrum (orange), intermodal phase differences (red) and normalized intensity  recontruction in time (blue). \textbf{a)} Computed GDD for the metasurface combined with different output couplers. \textbf{b)} \textit{ref-OC} in standard configuration with linear intermodal phase differences chirp. 37 ps at FWHM pulse emission. \textbf{c)} \textit{GTI87} in GTI configuration with flat intermodal phase differences. 6 ps at FWHM pulse emission. \textbf{d)} \textit{GTI62} in GTI configuration with flat intermodal phase differences and broader THz bandiwidth emission. 3.5 ps at FWHM pulse emission.}
    \label{fig:strong_inj}
\end{figure}

The limit in maximum QW comb THz bandwidth (BW) is not related to the power saturation of the injected RF modulation, but rather to a change in the operating regime of the VECSEL. Indeed, due to the limited dynamic range of the current ($I_{max} - I_{th}$) of the QC active material, at higher RF injection power ($\sim$ +35 dBm) even considering insertion losses, we expect that the RF current swing exceeds the dynamic range, and 
the lasing action is periodically switched on and off every cycle as the laser is pushed above NDR and the subbands become misaligned (see LIV curve in \ref{fig:fig1}(b)). 
The QC-VECSEL was re-tested in this strong modulation regime with \textit{ref-OC}; no signature of QW comb spectral map is observed, and the spectral broadening is asymmetrical with respect to the single-mode frequency although a broader spectrum (BW\textgreater 330 GHz) is measured. The DC spectrum product and the SWIFT spectrum are plotted together in Fig. \ref{fig:strong_inj}b, which exhibit good agreement indicating a high degree of coherence in the lasing emission. A linear chirp in the intermodal phase difference is also be observed. Consequently, the reconstructed time profile exhibits a periodic waveform with strong amplitude-modulation (AM) behavior with broad pulses (37 ps at full-width at half-maximum (FWHM)) and ps-wide sub-features, as visible in Fig. \ref{fig:strong_inj}(b).  This represents an actively mode-locked regime consistent with an on/off gain switching induced by the RF modulation of the optically-thin metasurface. 
\cite{barbieri2011coherent,hillbrand2020mode}. 
The dispersed intermodal phases prevent shorter pulse formation and suggest that dispersion compensation would be highly beneficial for producing shorter pulses in the amplitude-modulated regime.
To this end, the AM regime under strong RF modulation is further investigated using the \textit{GTI87} OC. The flat spectrum remains nearly unchanged (Fig. \ref{fig:strong_inj}c). However, thanks to the dispersion induced by the GTI configuration (Fig. \ref{fig:strong_inj}a), the linear chirp in the intermodal phase difference is corrected into a nearly flat phase profile. As a result, the pulse length at FWHM is reduced to 6 ps. 

A second dispersion-compensated output coupler \textit{GTI62} was specifically designed to provide matched dispersion compensation --- albeit imperfect --- around the metasurface resonance. 
Unlike \textit{GTI87}, which enhanced the parabolic effective  potential needed for the QW comb state, \textit{GTI62}  was designed to minimize total dispersion. Simulations estimate that the cavity GDD ranges from $-0.3$ to $0.3$~ps$^2$ over the lasing bandwidth in the \textit{ref-OC} case; this is reduced to values ranging from $-0.05$ to $0.2$~ps$^2$ for the \textit{GTI62} case (see Fir. \ref{fig:OCs_GDD}). 
 The emission bandwidth was increased to 600 GHz, of which $\sim$500 GHz was resolved with SWIFT spectroscopy. Under this condition, AM comb operation was observed where the FWHM pulse width was further reduced to 3.5 ps (Fig.\ref{fig:strong_inj}d).
In general, we observe that the effect of the dispersion compensation between the three OCs is quite significant,  as the initially broad, structured pulse narrows down by a factor of 10 to a single short actively-mode-locked pulse. 

\subsection*{Discussion and outlook}\label{conclusion}
We have conclusively shown that the THz QC-VECSEL platform can sustain optical frequency comb emission, as demonstrated through SWIFT spectroscopy characterization. Exploiting the lack of spatial hole burning in a QC-metasurface, this work extends the observation of quantum walk combs to non-ring cavities, provided that single-mode emission, an effective harmonic potential, and fast gain are maintained.
While a QC-VECSEL is larger than a monolithic waveguide-based QCL comb, the QC-VECSEL architecture also has some unique benefits crucial for THz sources. Specifically, we were able to modify the cavity dispersion through introduction of a variety of discrete GTI output couplers without the need for re-fabrication of the QC metasurface.
In one case, a GTI OC was used to restore a parabolic dispersion potential profile between optical modes, which increased in the quantum walk comb emission bandwidth and eliminated discontinuities in the expansion maps associated with TOD. In another case, a GTI OC was used for partial dispersion compensation; when combined with strong RF modulation, AM combs with flatter intermodal phase differences were obtained over more than 500 GHz. The result was actively mode locked QC-VECSELs with pulses as short at 3.5 ps. 
Overall, this work demonstrates that the THz QC-VECSEL is a highly flexible and practical platform for THz optical frequency combs, where the ability to switch between two operating regimes (FM and AM)\cite{hillbrand2020phase} is demonstrated by adjusting the RF and electrical bias in conjunction with dispersion engineering. 
Furthermore, the QC-VECSEL has appealing features for eventual dual comb spectroscopy (DCS) applications. For example, its tunable cavity length 
would allow convenient modification of the repetition rate (and thus control over the DCS spectral resolution and power-per-tooth). Also, QC-VECSELs exhibit excellent beam patterns and have scalable output power, which are important for any practical use. 
Finally, we believe there is considerable room for further improvement via gain and dispersion engineering, since QC metasurfaces are known to exhibit gain bandwidths exceeding 1 THz \cite{metasurface_coupled_resonator_Curwen_2020} , and the pulse durations observed here suggest that additional temporal compression remains accessible.

\section*{Methods}\label{sec4}
\subsection*{Sample layout and fabrication}


The output couplers (OCs) are fabricated by evaporation of an inductive Ti/Au metal-mesh on z-cut double-side-polished quartz substrates \cite{Mesh_coupler_Densing_1992,Multimode_Wu_2021}. The mesh dimensions (10 $\mu$m period and 4 $\mu$m metal linewidth) were chosen to give high reflectance (simulated $R \approx 99\%$) to maximize the possible lasing bandwidth. 
Fabrication is performed via a single photolithography step and evaporation of Ti/Au films (15/250 nm). In the ordinary configuration (referred to as ref-OC), the highly reflective mesh side of the OC is facing the metasurface, such that there is little field within the quartz substrate. In that configuration the quartz contributes negligible loss and dispersion, and the primary effect of the susbtrate is a slowly varying reflectance fringes due to residual etalon effects of the substrate. However, if the orientation of the OC is reversed, so that the quartz substrate faces the metasurface, the OC also acts as a Gires-Tournois interferometer (GTI), which contributes additional dispersion associated with the standing wave in the quartz substrate. Two GTI-OCs were fabricated with the same mesh on quartz substrates polished to custom thicknesses: 87 $\mu$m and 62 $\mu$m (referred to as \textit{GTI87} and \textit{GTI62}).  Full-wave COMSOL electromagnetic simulations provide the  reflectance, absorbance, and group delay dispersion (GDD) of the GTI-OC including the frequency-dependent permittivity of quartz \cite{optical_constants_Loewenstein_1973} and Drude-model for the Au mesh (see Supplementary Material). The  reflection coefficient phase $\theta$ is simulated using the $e^{j\omega t}$ convention, and the GDD is given by $GDD = -\frac{d^{2}\theta}{d\omega^{2}}$. The final substrate thicknesses were chosen such that the regions of highest reflectance overlap with the desired metasurface resonance, and to add to or subtract from the total cavity GDD. 

The intra-cryostat VECSEL cavity is constructed using an off-axis-paraboloid mirror with a one-half inch focal length. Alignment takes place using a visible laser focusing to the center of the metasurface. The reflected diffraction pattern originating from the metasurface periodicity is then aligned to the diffraction pattern associated with the OC mesh. This is the case regardless of whether the OC is in a GTI orientation or typical orientation (mesh facing the metasurface). Once the metasurface is aligned to the OAP mirror, the OC can be swapped with another by repeating the second alignment step. 

\subsection*{LIV measurement}
The device is cooled down at 25 K by using a liquid helium transfer line. The DC bias is applied via Keithley 2400 Sourcemeter. The collimated THz beam passes through a TPX windwow (transparent to THz) and it is detected by a calibreted power meter (TK Terahertz Absolute Power Meter System).
\subsection*{Spectral map}
The spectral maps are recorded through a Fourier Transform Infrared spectrometer (FTIR) Vertex 80v (BRUKER) equipped with a room-temperature deuterated triglycine sulfate (DTGS) detector.
\subsection*{RF modulation}
The RF signal is generated using a Rohde\&Schwarz SMB 100A microwave signal generator and amplified with a MiniCircuits ZVE-3W-183+ microwave amplifier (+35 dBm). The DC bias and the RF modulation are applied at the same port of the PCB via biastee to protect the amplifier.
\begin{itemize}
\item FM standard configuration (reported in Fig.\ref{fig:QW_ref_OC}a): I = 168 mA, f\textsubscript{rep} = 6.3864 GHz, P = +7 dBm.
\item QW comb laser standard configuration (reported in Fig.\ref{fig:QW_ref_OC}b): I = 160 mA, f\textsubscript{rep} = 6.391 GHz, P = +17 dBm.
\item QW comb laser GTI87 (reported in Fig.\ref{fig:fig3}): I = 175 mA, f\textsubscript{rep} = 5.929 GHz, P = +19 dBm.
\item AM standard configuration (reported in Fig.\ref{fig:strong_inj}b): I = 145 mA, f\textsubscript{rep} = 6.387 GHz, P = +35 dBm.
\item AM GTI87 (reported in Fig.\ref{fig:strong_inj}c): I = 160 mA, f\textsubscript{rep} = 5.928 GHz, P = +35 dBm.
\item AM GTI62 (reported in Fig.\ref{fig:strong_inj}d): I = 150 mA, f\textsubscript{rep} = 5.822 GHz, P = +35 dBm.
\end{itemize}
\subsection*{SWIFT measurements}
The room-temperature Schottky mixer is incorporated in the FTIR setup and is used as a fast detector. The lasing optical beatnote is then IQ demodulated at the injected RF frequency by a Rohde\&Schwarz FSW67 RF spectrum analyzer.

\section*{Acknowledgements}

M.R. thanks A. Forrer for his earlier work on the SWIFTS setup and analysis code. M.R. thanks V. Digiorgio and I. Heckelmann for fruitful discussions.

Financial support from MINT SNF grant n 10000934  is gratefully acknowledged. 
This work was part of 23FUN03 COMOMET that has received funding from the European Partnership on Metrology, co-financed by the European Union’s Horizon Europe Research and Innovation Programme and from by the Participating States. Funder ID: 10.13039/100019599  . This work partially funded by the National Science Foundation grant 2440163.This work was performed, in part, at the Center for Integrated Nanotechnologies, an Office of Science User Facility operated for the U.S. Department of Energy (DOE) Office of Science. Sandia
National Laboratories is a multimission laboratory managed and operated by National Technology
and Engineering Solution of Sandia, LLC., a wholly owned subsidiary of Honeywell International,
Inc., for the U.S. Department of Energy’s National Nuclear Security Administration under contract
DE-NA-0003525.

\bibliography{bib_RAFFA,bib_Digio,bib_ThesisUrban,Biblio_Giak}
\bibliographystyle{naturemag_noUR}

\end{document}





\maketitle 
\tableofcontents

\section{Harmonic and sub-harmonic RF driving for standard configuration}
\begin{figure}
    \centering
    \includegraphics[width=1\linewidth]{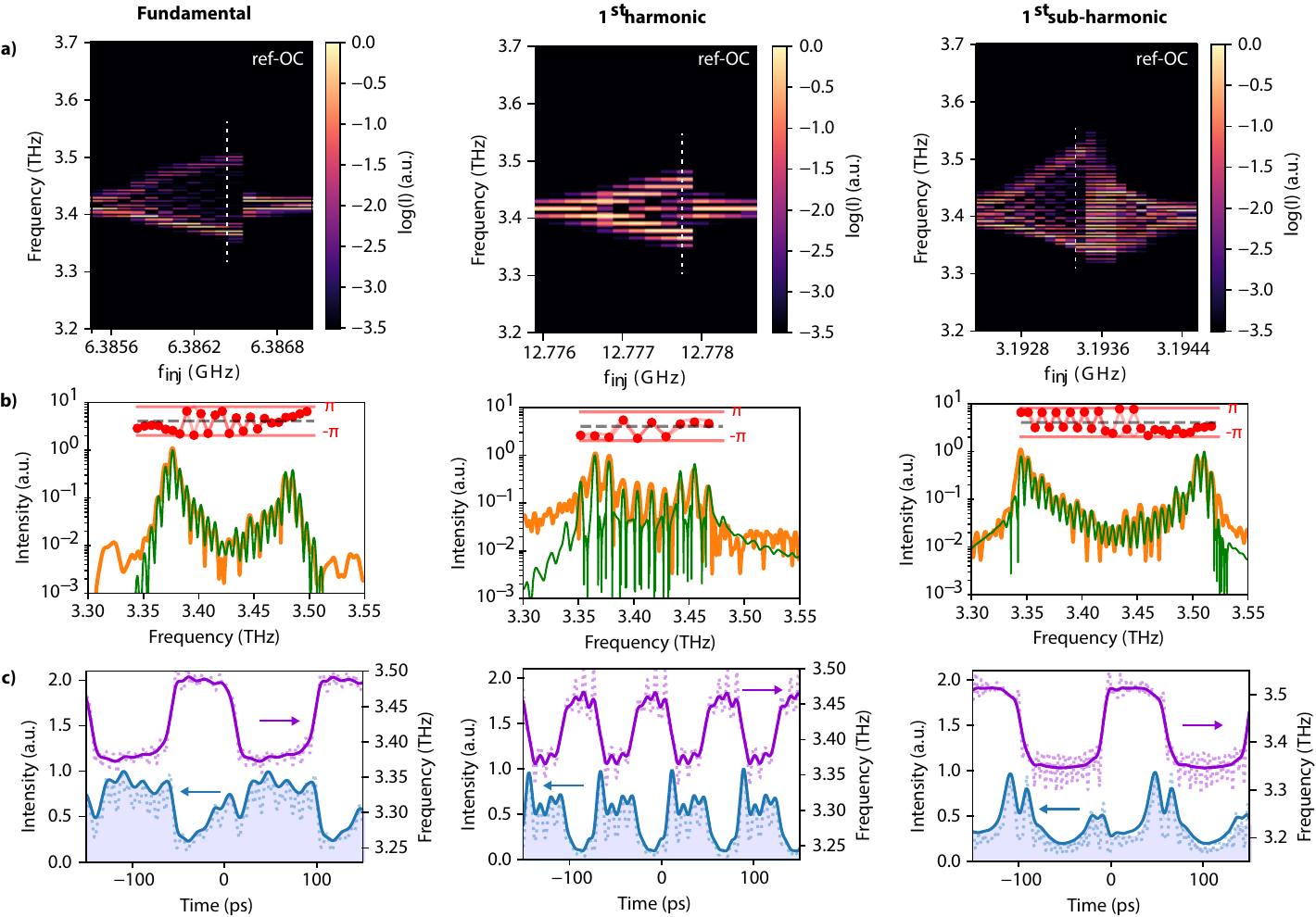}
    \caption{\footnotesize \textbf{RF driving at fundamental, 1\textsuperscript{st} harmonic and 1\textsuperscript{st} sub- harmonic of the cavity resonant frequency.} \textbf{a)} Spectral maps domintaed my third-order dispersion with \textit{ref-OC}. \textbf{b)} Spectral product (green), SWIFT spectrum (orange) and intermodal phase differences (red). \textbf{c)} Normalized intensity (blue) and instantaneous frequency (violet) reconstructed [dots] and smoothed [solid trace] over time by SWIFT measurements.}
    \label{fig:summary_3rd_order}
\end{figure}
In Fig.\ref{fig:summary_3rd_order}a, spectral maps injecting RF modulation at fundamental, 1\textsuperscript{st} harmonic and 1\textsuperscript{st} sub-harmonic are plotted. In the following panel b) and c), SWIFT results close to resonance are presented: optical frequency comb emission is preserved in all the injection regimes, with signature jump in spectral maps and double discontinuity point in instantaneous frequency. Even-harmonic spectral emission was achieved when the frequencies were injected near twice the round-trip frequency. Conversely, when an RF signal at half the round-trip frequency was fed into the device, the strong nonlinearity of the QC active material locked the optical modes at the fundamental frequency (as reported in Ref. 29), enabling the observation of a FM comb state under all injection conditions.

\section{Numerical simulations}
\begin{figure}
    \centering
    \includegraphics[width=0.8\linewidth]{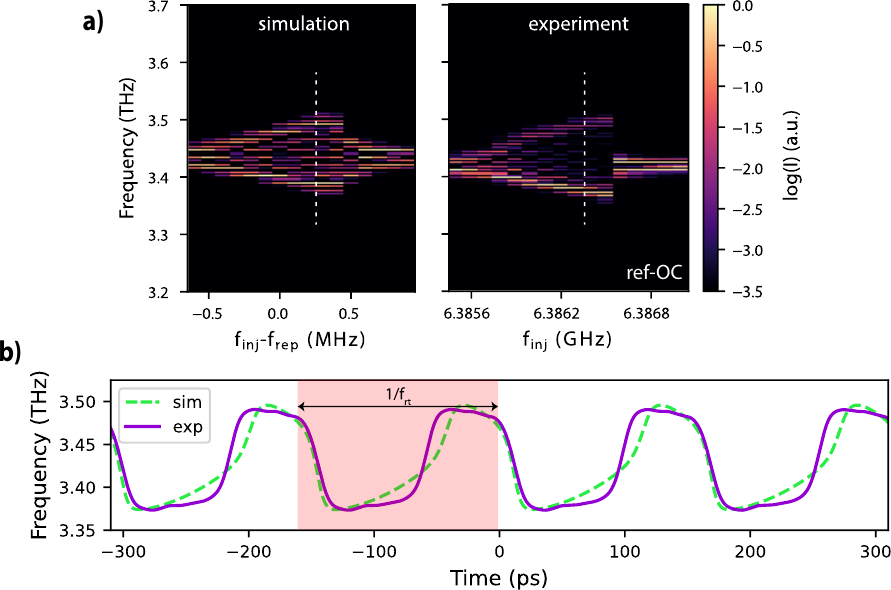}
    \caption{\footnotesize \textbf{Third-order dominated FM comb regime.} \textbf{a)} Spectral maps in simulation and experiment. \textbf{b)} Double discontinuity of the instantaneous frequency over time in one cavity round-trip close to resonance due to the third order dispersion in simulation and experiment.}
    \label{fig:sim}
\end{figure}

The dynamics of QW comb formation are simulated by solving the complex Ginzburg–Landau equation, a widely used nonlinear differential equation for modeling extended nonequilibrium systems:
\begin{equation}
    T_R\frac{\partial F}{\partial T} = (  [g- \alpha] -g\frac{I}{I_{sat}})F + (D_g +iD_i)\frac{\partial^2 F}{\partial t^2} +iMcos(\Omega_{res}t)F
\end{equation}
The term on the left-hand side describes the slow-time evolution of the pulse envelope over many cavity round trips, where $T_R$ is the cavity round-trip time. The terms linear in $F$ account for the net gain ($g-\alpha$), the gain-saturation contribution governed by the instantaneous intensity $I=|F|^2$, and the phase modulation of the RF signal with amplitude $M$at the cavity resonance frequency $\Omega_{res}$.The gain curvature $D_g$ and dispersion $D_i$ govern variations in the pulse envelope on the fast timescale $t$, within a single cavity round trip.

A split-step Fourier method is used to compute, in small steps, the evolution of a single mode into a QW comb under RF modulation. At each step, the intracavity field is first transformed into the spectral domain, where the linear operator accounting for dispersion and gain curvature is applied. The field is then transformed back into the cavity-coordinate domain, where the local operator is applied. This operator includes saturated gain, losses, and RF-driven phase modulation. In particular, the gain-saturation term is evaluated in the local domain because it depends on the instantaneous field intensity, $|F|^2$. The resulting field is used as the initial condition for the next step, and the procedure is repeated until the spectrum stabilizes.

In Fig.\ref{fig:sim}, results of the simulations, including anharmonicity of the potential in the synthetic lattice, are reported. The impact of the third-order term in the dispersion is well reproduced by the simulations: both the abrupt jump and the double discontinuity points in instantaneous frequency in time match the experimental observation.
\section{Bistability}

\begin{figure}
    \centering
    \includegraphics[width=0.8\linewidth]{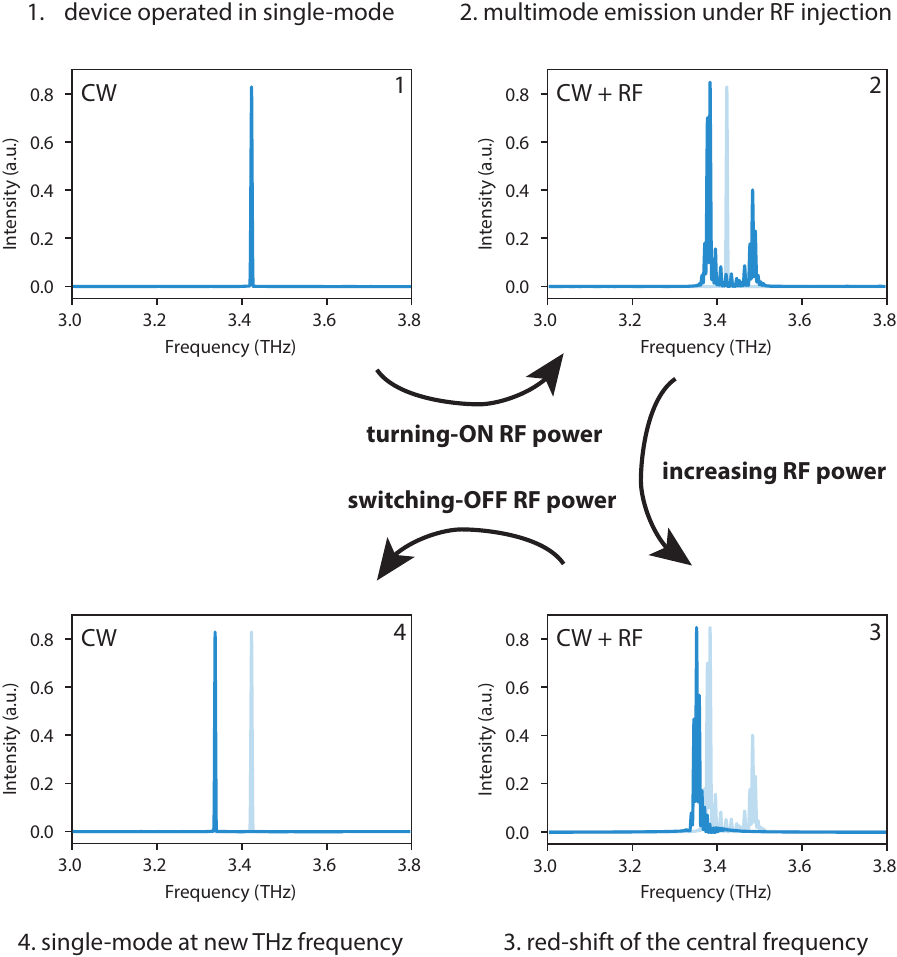}
    \caption{\footnotesize \textbf{Bistable evolution under RF modulation.}}
    \label{fig:bistability}
\end{figure}

\begin{figure}
    \centering
    \includegraphics[width=1\linewidth]{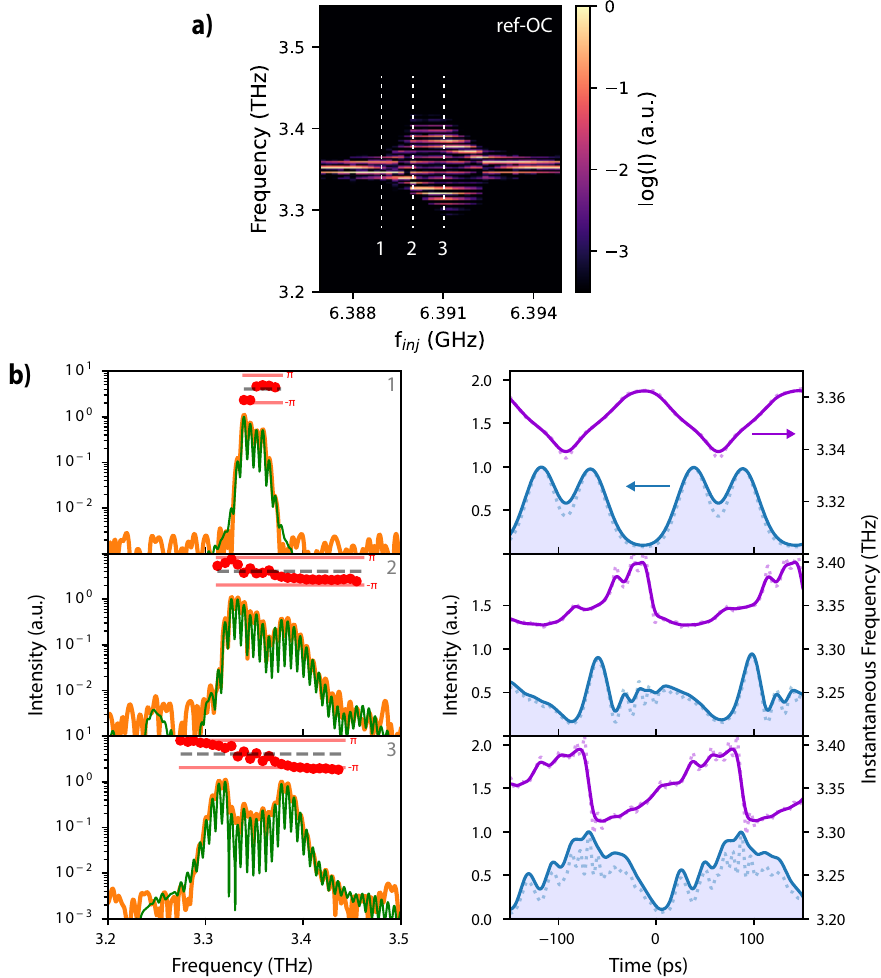}
    \caption{\footnotesize \textbf{Summary of SWIFT measurements after exploiting bistable single mode frequency-shift}. Off-resonance \textbf{(1)}, slightly shifted-from-resonance \textbf{(2)} and on-resonance \textbf{(3)}.}
    \label{fig:QW_refOC}
\end{figure}

\begin{figure}[h]
  \centering
  \begin{minipage}{0.45\textwidth}
    \centering
    \includegraphics[width=\linewidth]{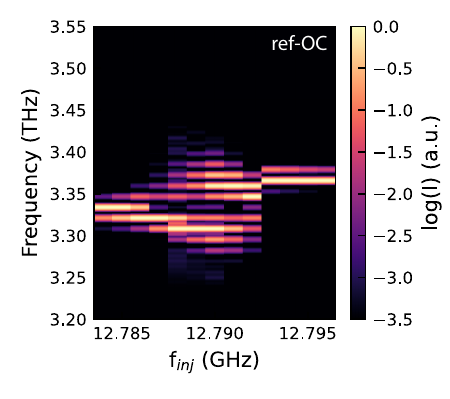}
  \end{minipage}
  \begin{minipage}{0.45\textwidth}
    \centering
    \includegraphics[width=\linewidth]{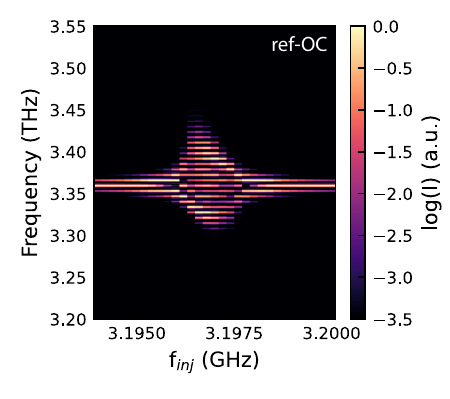}
  \end{minipage}
  \caption{\footnotesize \textbf{QW comb spectral map after exploiting bistable single mode frequency-shift.}  1\textsuperscript{st} harmonic (\textbf{left}) and 1\textsuperscript{st} sub-harmonic (\textbf{right}).}
  \label{QW_harmonic_subharmonic}
\end{figure}

As reported in Extended Data Fig.3b, bistability is exploited to shift the THz frequency of the "walker" and to obtain a proper QW comb spectral map. In Fig.\ref{fig:bistability} the evolution of the lasing emission is described:
\begin{enumerate}
    \item the device is operated in CW and exhibits single-mode lasing at 3.42 THz;
    \item RF modulation is applied at fundamental resonant cavity frequency with +7 dBm at the source;
    \item RF power is further increased to +18 dBm at the source to appreciate the abrupt red-shift of the central frequency;
    \item RF power is turned-OFF and the single-mode emission is then centered at 3.35 THz, until a bias reset of the device.
\end{enumerate}
Following the same procedure described in the Main text, proper QW comb spectral maps are reported for both 1\textsuperscript{st} harmonic and 1\textsuperscript{st} sub-harmonic RF modulation in Fig.\ref{QW_harmonic_subharmonic}.

\section{GTI configuration}
\begin{figure}[h]
    \centering
    \includegraphics[width=1\linewidth]{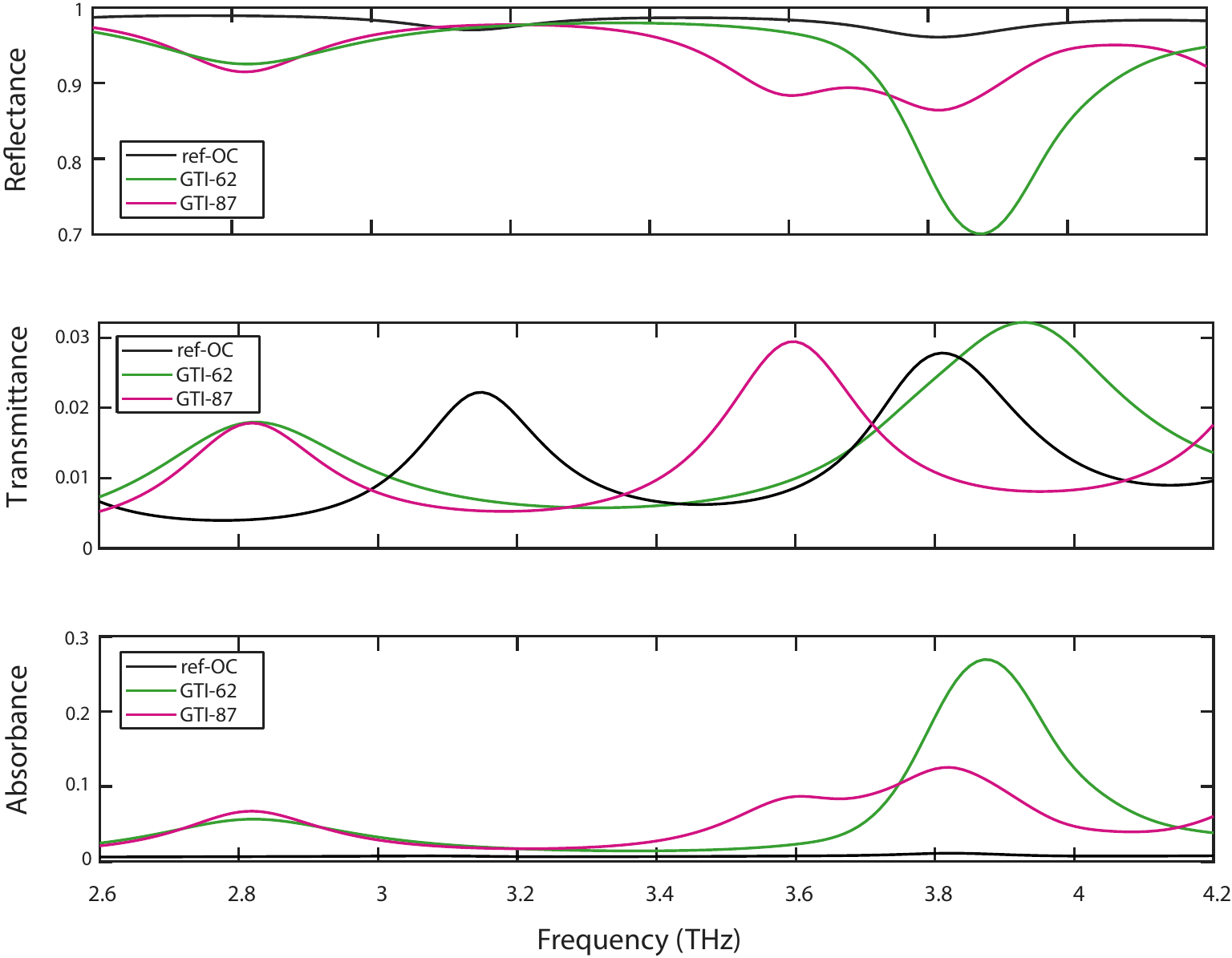}
    \caption{\footnotesize \textbf{Simulations of the optical properties of each Output Couplers.}}
    \label{fig:RTA}
\end{figure}
\begin{figure}[h]
    \centering
    \includegraphics[width=0.7\linewidth]{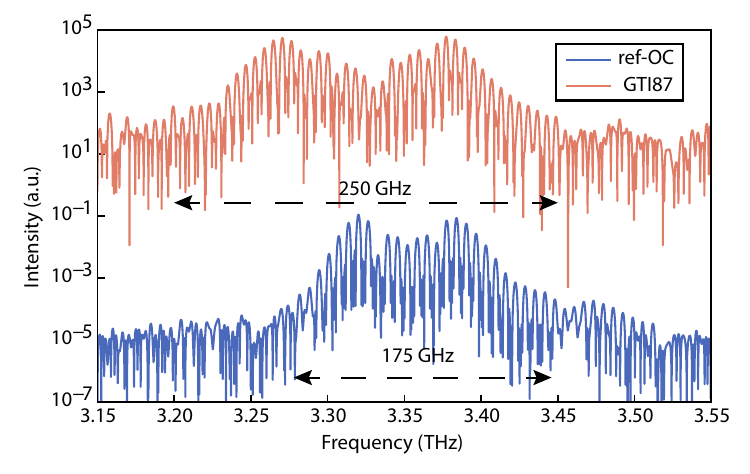}
    \caption{\footnotesize \textbf{QW comb spectra at resonance for standard configuration \textit{ref-OC} and GTI configuration \textit{GTI87}.}}
    \label{fig:spectra_QW_comparison}
\end{figure}

\begin{figure}[h]
    \centering
    \includegraphics[width=1\linewidth]{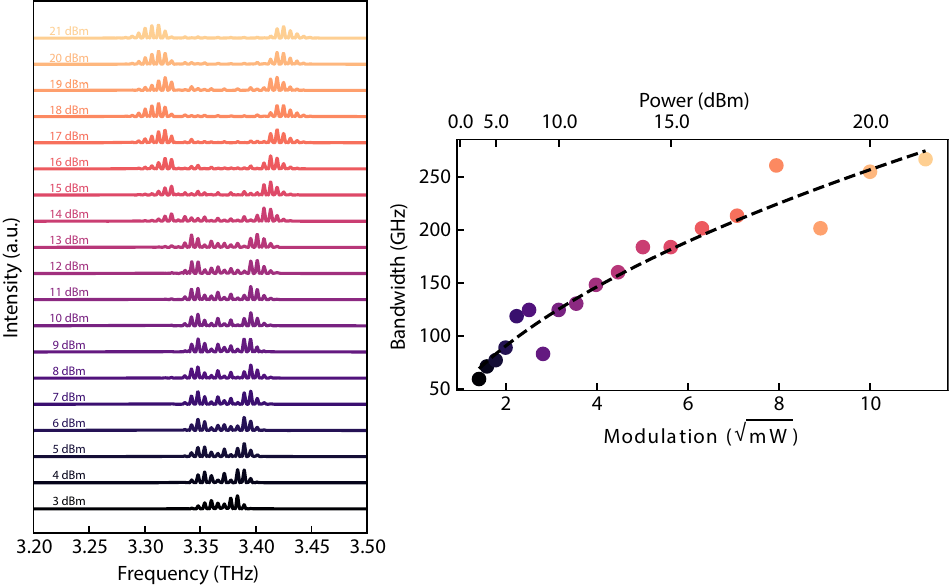}
    \caption{\footnotesize \textbf{QW comb RF power sweep with \textit{GTI87}.}}
    \label{fig:QW_power_sweep}
\end{figure}
\begin{figure}[h]
    \centering
    \includegraphics[width=0.9\linewidth]{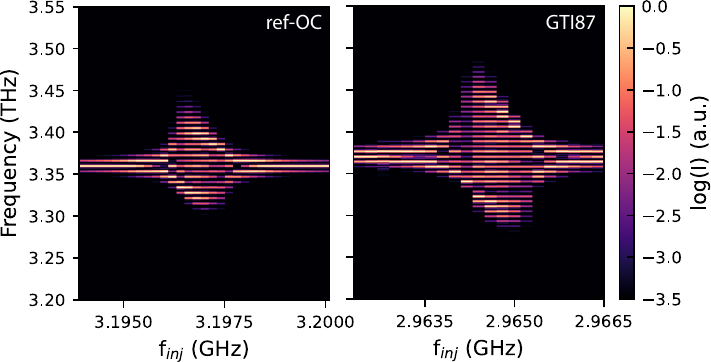}
    \caption{\footnotesize \textbf{Comparison QW comb spectral maps at subharmonic RF modulation}. \textit{ref-OC} (\textbf{left}) and \textit{GTI87} (\textbf{right}).}
    \label{fig:QW_sub_comp}
\end{figure}

The metasurface reflection coefficient spectrum exhibits a nearly $2\pi$ phase shift across its Lorentzian-type resonance; 
when the metasurface is at design bias, the intersubband transitions in the QC material contribute little to dispersion. This suggests the dispersion is relatively independent of laser bias, and thus suitable for compensation using external elements. A natural first choice to consider is the output coupler mirror, which is typically fabricated by patterning a quartz or high-resistivity silicon substrate with an inductive metal mesh. 
The reflectance of the output coupler mirror is controlled by the mesh dimensions. Output couplers are designed for high reflectance (95-99\%), and provided that the metal mesh faces the QC metasurface, contribute negligible dispersion across the gain bandwidth of a metasurface. However, if the metal mesh faces away from the QC-metasurface, the Fabry–Pérot resonances in the substrate contribute an additional phase shift, and the OC acts also as a Gires Tournois interferometer (GTI). Hence, the thickness of the substrate can be tuned to add more or less dispersion to the cavity at different frequencies. The reflectance, transmittance and absorbance of \textit{ref-OC}, \textit{GTI87} and \textit{GTI62} are reported in Fig. \ref{fig:RTA}. Fig. \ref{fig:spectra_QW_comparison} compares the QW comb laser spectra at resonance for the standard (\textit{ref-OC}) and GTI configurations (\textit{GTI87}). The restored harmonic potential made possible to increase the THz bandwidth from 175 GHz to 250 GHz. QW comb laser spectra at resonant frequency as function of the injected RF power are shown in Fig.\ref{fig:QW_power_sweep}. The bandwidth of the THz emission is fit with P\textsuperscript{1/4} dependence as predicted by QW comb laser theory.

The benefit of the GTI output couplers in terms of effective potential is visible comparing the expansion maps at sub-harmonic RF modulation in Fig. \ref{fig:QW_sub_comp}. Even in this case, the BW at resonant driving reaches 250 GHz in the dispersion engineered configuration.

\section{Strong RF modulation maps}
\begin{figure}[h]
    \centering
    \includegraphics[width=\linewidth]{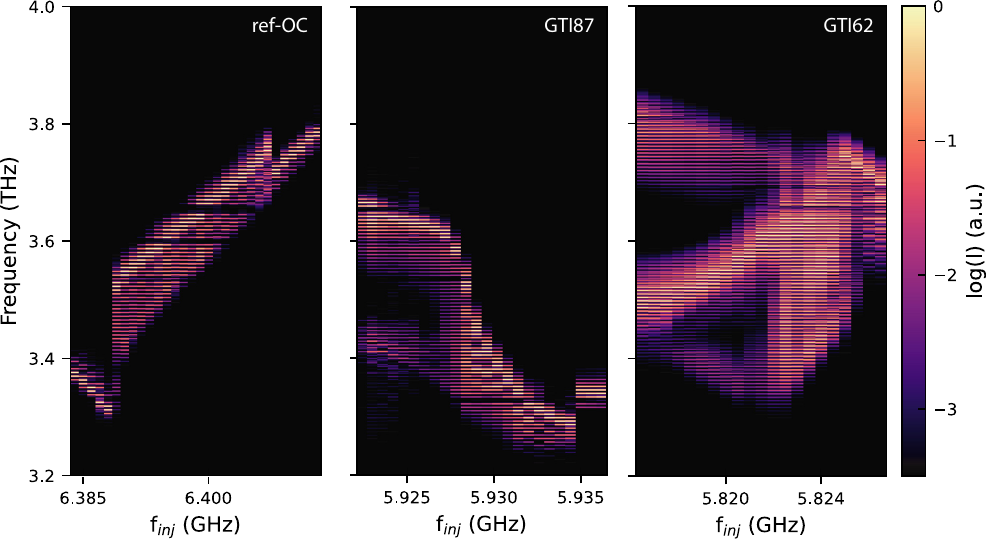}
    \caption{\footnotesize \textbf{Strong RF modulation maps}. (\textbf{left}) ref-OC (\textbf{center}) GTI87 (\textbf{right}) GTI62}
    \label{fig:strong_maps}
\end{figure}

As described in the Extended Data, under strong RF injection the QW comb spectral maps are not retrieved. In Fig. \ref{fig:strong_maps} the asymmetric broadening as function of the injected RF frequency is shown for each output coupler configuration.